\documentclass[aps,prb,preprint,superscriptaddress]{revtex4-1}
\usepackage{amsmath}
\usepackage{amssymb}
\usepackage{graphicx}   
\usepackage{mhchem}
\usepackage{amsfonts}
\usepackage{adjustbox}
\usepackage{booktabs}

\begin{document}

\newcommand{\degrees}{$^\circ$C}
\newcommand{\nbcpo}{Na$_{2}$BaCo(PO$_{4}$)$_{2}$}
\newcommand{\nbmpo}{Na$_{2}$BaMg(PO$_{4}$)$_{2}$}

\title{Strong quantum fluctuations in a quantum spin liquid candidate with a Co-based triangular lattice}

\author{Ruidan Zhong}
\author{Shu Guo} 
\affiliation{Department of Chemistry, Princeton University, Princeton NJ 08544, USA}
\author{Guangyong Xu}
\affiliation{NIST Center for Neutron Research, National Institute of Standards and Technology, Gaithersburg MD 20899, USA}
\author{Zhijun Xu}
\affiliation{NIST Center for Neutron Research, National Institute of Standards and Technology, Gaithersburg MD 20899, USA}
\affiliation{Department of Materials Science and Engineering, University of Maryland, College Park, MD 20742, USA}
\author{Robert J. Cava}
\affiliation{Department of Chemistry, Princeton University, Princeton NJ 08544, USA}

\date{\today}

\begin{abstract}
Currently under active study in condensed matter physics, both theoretically and experimentally, are quantum spin liquid (QSL) states, in which no long-range magnetic ordering appears at low temperatures due to strong quantum fluctuations of the magnetic moments. The existing QSL candidates all have their intrinsic disadvantages, however, and solid evidence for quantum fluctuations is scarce. Here we report a new compound, \nbcpo, a geometrically frustrated system with effective spin-1/2 local moments for Co$^{2+}$ ions on an isotropic two-dimensional triangular lattice. Magnetic susceptibility and neutron scattering experiments show no magnetic ordering down to 0.05 K.  Thermodynamic measurements show that there is a tremendous amount of magnetic entropy present below 1 K in zero applied magnetic field. The presence of localized low-energy spin fluctuations is revealed by inelastic neutron measurements. At low applied fields, these spin excitations are confined to low energy and contribute to the anomalously large specific heat. In larger applied fields, the system reverts to normal behavior as evident by both neutron and thermodynamic results. Our experimental characterization thus reveals that this new material is an excellent candidate for the experimental realization of a quantum spin liquid state.

\end{abstract}

\maketitle
\section*{Significance statement}
The experimental search and verification of theoretically predicted quantum spin liquids (QSLs) is challenging, especially because there is no universally accepted evidence to support such exotic quantum states. At present, broad continuous magnetic excitations, i.e. a spinon continuum, observed via inelastic neutron scattering, is considered as the most crucial evidence for a QSL, as described in many studies on the existing QSL candidates. However, as another potential origin of such broad spin excitations, disorder effects in those candidates cannot be neglected. Here we present a new compound with a Co-based triangular lattice that is structurally perfect and without intrinsic disorder. We present experimental results that suggest that this new compound is an ideal QSL candidate for future studies. 

\section*{Introduction}

A quantum spin liquid (QSL) is a unique type of spin liquid where strong quantum fluctuations prevent spins from ordering, which thus remain in a disordered liquid-like state even near absolute zero \cite{Mila2000, Balents2010}. Previous studies have shown that QSL states tend to emerge in low-spin geometrically frustrated systems, in which the interactions among the limited magnetic degrees of freedom are restricted by crystal geometry. This leads to a strong enhancement of quantum fluctuations \cite{Moessner2006}. Several compounds with geometrically frustrated lattices have been proposed as QSL candidates \cite{Yamashita2008, Yamashita2010, Han2012, Shen2016, Paddison2017}. At present, the most widely accepted evidence for the QSL state is a broad continuous magnetic excitation observed in inelastic neutron scattering (INS) \cite{Han2012, Shen2016, Paddison2017} believed to be associated with the fractionalized quasiparticles, spinons, that are expected for a QSL state \cite{Balents2010}. 
Structural disorder and anisotropic spin interactions due to site-mixing or lattice distortions in QSL candidates often result in a spin glass or magnetically ordered ground state, however, which can give rise to an impostor spin-liquid-like state \cite{Zhu2017} and produce a broad continuum in INS \cite{Ma2018}. Unfortunately, the problem of site-mixing or lattice distortion in the existing QSL candidates cannot be neglected \cite{Freedman2010, Han2012prl, Li2015}. Similar fractionalized excitations have also been seen in Cs$_{2}$CuCl$_{4}$ \cite{Coldea2001} and Ba$_{3}$CoSb$_{2}$O$_{9}$ \cite{Zhou2012, Ito2017} but those materials are known to have magnetically ordered states at low temperature \cite{Zhou2012, Coldea1996}. Thus all of the above materials evoke skepticism about whether the observation of a broad continuum feature in the INS can be uniquely attributed to the QSL state. Thus when looking for a more robust characterization of the QSL state, a new ideal candidate with no disorder, isotropic magnetic exchange coupling, and a well-defined crystal structure is a necessary but challenging goal to pursue.

Here we propose a new quantum spin liquid compound, \nbcpo, in which Co$^{2+}$ with effective $S$=1/2 resides on a geometrically frustrated triangular lattice. A significant advantage of this material is that it does not display any significant site-mixing or lattice distortions -- the magnetic Co$^{2+}$ ions constitute isotropic triangular lattice layers with simple packing. Further, single crystals can be grown. \textit{dc} magnetic susceptibility and elastic neutron scattering experiments reveal no long-range magnetic ordering down to 0.3 K in the presence of antiferromagnetic interactions on the order of 32 K. In addition, \textit{ac} magnetic susceptibility measured down to 0.05 K indicates no spin glass transition. The magnetic specific heat on a single crystal reveals the presence of a large magnetic entropy at low temperature. INS measurements indicate the presence of strong localized magnetic fluctuations, which are nearly temperature-independent and have exotic behavior under a magnetic field. These observations indicate that this new material is an excellent QSL candidate. 

\section*{Results}

\begin{figure*}[t]
\centering
\includegraphics[width=13cm,height=9.0cm]{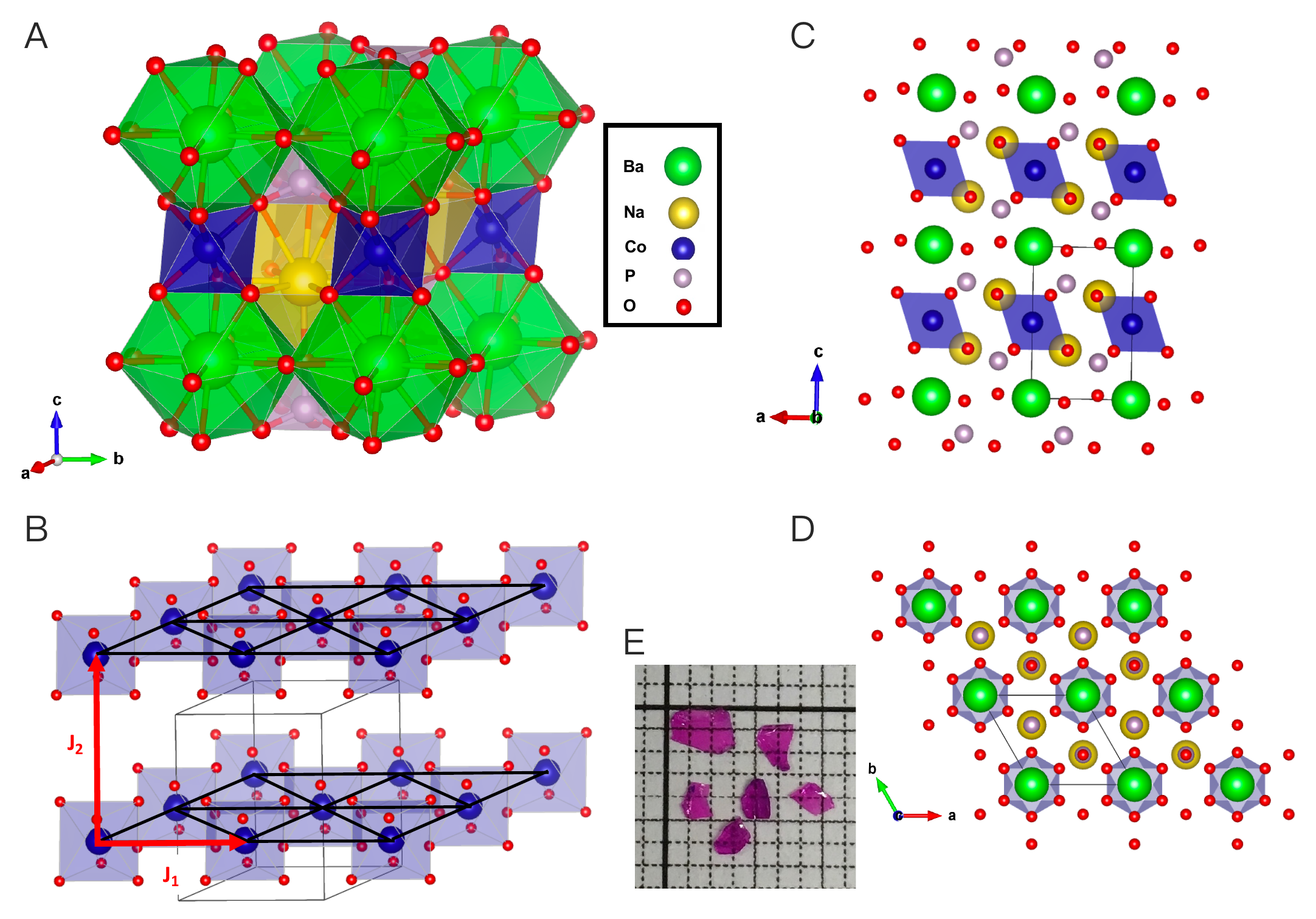}
\caption{The crystal structure of \nbcpo. \textit{(A)} The crystal structure for \nbcpo based on the packing of M-O polyhedra. \textit{(B)} The triangular layer of CoO$_{6}$ octahedra in the $ab$ plane, with $J_{1}$ and $J_{2}$ marked indicating the super-exchange coupling between the nearest neighbor Co$^{2+}$ ions in the same plane and the neighboring plane. \textit{(C)} view emphasizing the layers of CoO$_{6}$ octahedra along the $c$-axis viewed from the b-axis. \textit{(D)} Top view of the relative orientation of the CoO$_{6}$ and other groups in the 2D triangular layers. The dashed lines in \textit{(C)} and \textit{(D)} indicate the unit cell. \textit{(E)} Photo of the pink single crystals of \nbcpo.}\label{fig:fig1}
\end{figure*}

Pink single crystals of \nbcpo\ harvested from flux method growths have a layered morphology with hexagonal faces, shown in Figure~\ref{fig:fig1}\textit{E}. Single-crystal X-ray diffraction (XRD) data for \nbcpo\ indicate a trigonal crystal structure, space group P$\bar{3}$m1, with lattice constants of $a$ = $b$ = 5.3185(1) {\AA} and $c$=7.0081(1) {\AA}. Detailed crystal data and crystal structure refinement results are presented in Tables 1\&2. A similar structure has been reported for several Na$_{2}$Ba$M$(VO$_{4}$)$_{2}$ ($M$ = Ni, Co, Mn, and Fe) compounds \cite{Reub2018, Nakayama2013}, while the \nbcpo\ compound studied here is reported for the first time. Because of the distinct difference in the constituent atoms’ charge and ionic size, disorder due to the site-mixing is a-priori expected to be negligible, as is shown to be the case in the structural refinements (Table 2). Further, the single crystal refinement indicates the material is stoichiometric, with a simple whole number ratio of constituents. Schematic plots of the crystal structure are shown in Fig.~\ref{fig:fig1}\textit{A, C}, and \textit{D}. Triangular layers of magnetic CoO$_{6}$ octahedra are found in a simple A-A-A stacking pattern; the lattice disruptions due to stacking faults frequently seen in other hexagonal layered geometrically frustrated magnets are not seen in this material because there are no van der Waals-bonded planes present. The magnetic layers are instead separated by a single layer of nonmagnetic BaO$_{12}$ polyhedra, with [PO$_{4}$]$^{3-}$ units and Na$^{+}$ filling the gaps in the cobalt oxide layers. Since Co is a 3$d$ transition element, the magnetic exchange coupling strength depends on the overlap of metal-oxygen and oxygen-oxygen bonds (i.e. superexchange). As a result, the coupling between the nearest neighbor (NN) Co$^{2+}$ ions in the same triangular layer (Fig.~\ref{fig:fig1}\textit{B}, $J_{1}$) is through a Co$-$O$-$O$-$Co super-superexchange path, while the exchange coupling with Co$^{2+}$ ions in the neighboring layers ($J_{2}$) is through Co$-$O$-$O$-$O$-$Co super-super-superexchange. Therefore, despite the fact that the Co$-$Co spacing between interlayers (7.00 {\AA}) and intralayer (5.32 {\AA}) are moderately close, this is deceptive. $J_{2}$, the coupling between planes, should be much smaller than $J_{1}$, the in-plane exchange, due to the weaker exchange path. Therefore, the system can be considered as an effectively two-dimensional (2D) triangular magnet, and thus may show the effects of strong geometrical frustration.  

\begin{figure*}[t]
\centering
\includegraphics[width=.5\linewidth]{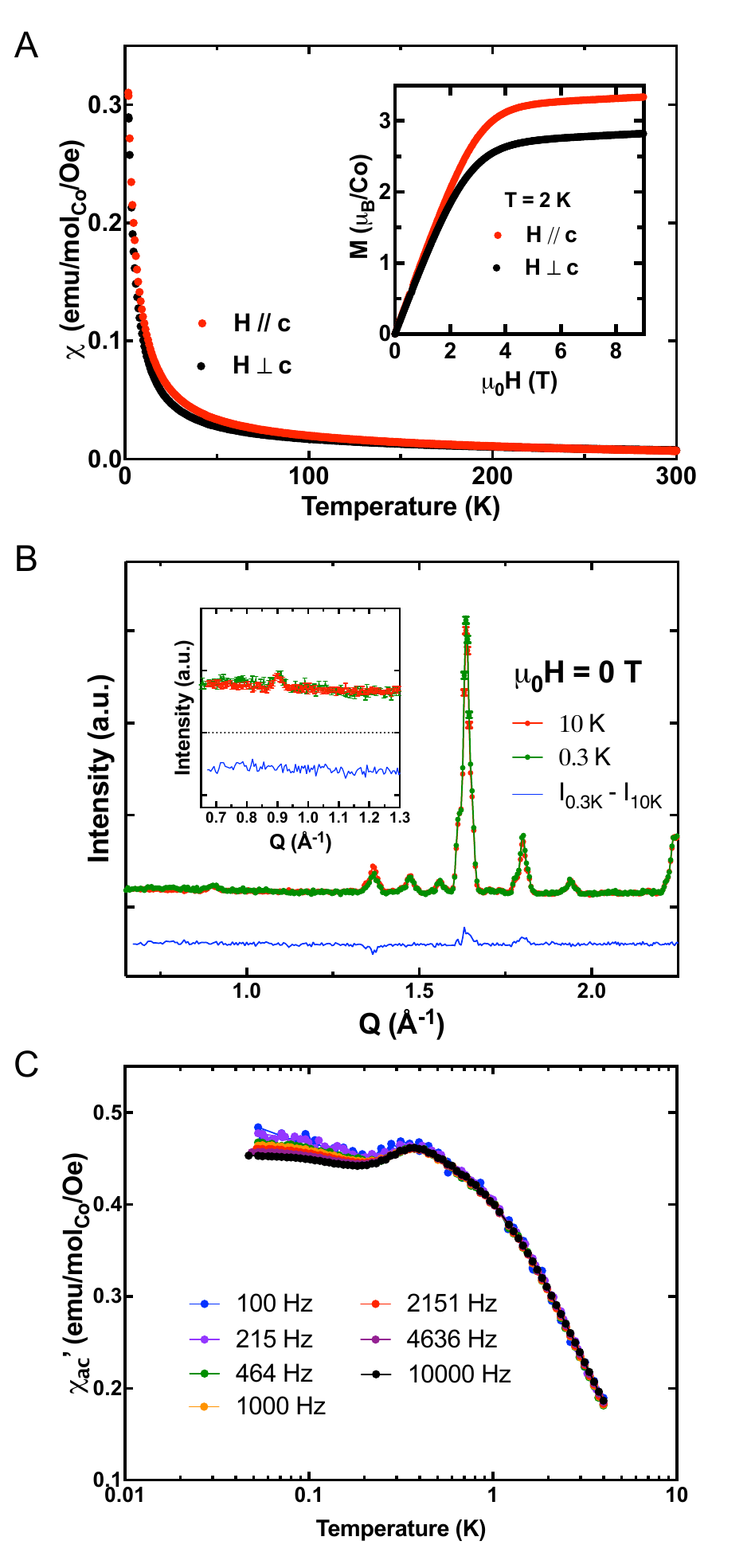}
\caption{No long-range or glassy magnetic ordering is observed in \nbcpo\ down to 0.05 K. \textit{(A)} Temperature dependence of the \textit{dc} susceptibility down to 1.8 K measured with $\mu_{0}H$ = 0.1 T on a single crystal sample oriented with the external field perpendicular to the c-axis (H $\perp$ c, black) and parallel to the c-axis (H $\parallel$ c, red). Inset: Field dependence of magnetization measured in both crystal orientations. \textit{(B)} Neutron powder diffraction pattern for \nbcpo\ measured at 0.3 K (green) and 10 K (red), as well as their difference (blue). Inset: Zoomed-in view of neutron diffraction pattern showing the low $Q$ region. \textit{(C)} Temperature dependence of the real part of the \textit{ac} magnetic susceptibility measured down to 0.05 K. }\label{fig:fig2}
\end{figure*}

Figure~\ref{fig:fig2}\textit{A} shows the dc magnetic susceptibility of \nbcpo\ down to 1.8 K. In contrast to the analogous materials Na$_{2}$BaCo(VO$_{4}$)$_{2}$ and Na$_{2}$BaNi(VO$_{4}$)$_{2}$, which have ordering transitions at 3.9 K and 8.4 K, respectively \cite{Nakayama2013}, no magnetic phase transition is observed for \nbcpo. Since Co$^{2+}$ is a Kramers ion, the effective magnetic moment of Co$^{2+}$ can be regarded as $S$=1/2 \cite{Shirata2012, Lines1963}, similar to what is observed in isostructural Na$_{2}$BaCo(VO$_{4}$)$_{2}$ \cite{Nakayama2013}. The Curie-Weiss law was applied to fit the susceptibility from 200 K to 300 K without including a Pauli susceptibility $\chi_{0}$. From the fits, we find $\Theta_{CW, \perp}$ = -31.9 K, $\mu_{eff,\perp}$ = 4.96 $\mu_{B}$, and $\Theta_{CW, \parallel}$ = -32.6 K,$\mu_{eff,\parallel}$ = 5.87 $\mu_{B}$ for the magnetic field applied perpendicular and parallel to the $c$-axis, respectively, indicating that the antiferromagnetic interactions dominate. The large effective moment suggests a nonnegligible orbital contribution to the magnetic moment. Using these fitted $\Theta_{CW}$ values, we estimate the exchange interaction constant $J/k_{B}$ to be 21.4 K \cite{Ma2018, Li2015prl}. Despite the expected difference in the exchange coupling based on the crystal structure, the magnetic anisotropy is significant but not large, further confirmed by the field-dependent magnetization shown in the inset of Figure~\ref{fig:fig2}\textit{A}. Due to the nearly isotropic magnetization in response to field, the sample orientation under field is only a minor factor in the subsequent physical measurements.

Neutron powder diffraction data obtained at 0.3 K and 10~K provide more information about the magnetic characterization of the new compound. From the above dc susceptibility results, the system is far from ordering at 10 K. By comparing the neutron diffraction peaks measured at both temperatures (Fig.~\ref{fig:fig2}\textit{B}) we find no extra superlattice peaks in the 0.3 K pattern compared to the one measured at higher temperature, and the intensities differ only within error. The magnetic peaks associated with long-range magnetic ordering should be strong at small $Q$ due to the magnetic form factor. Thus, the blow-up shown in the inset of Fig.~\ref{fig:fig2}\textit{B} further indicates there is no magnetic ordering\cite{Zhou2012} down to 0.3 K. No frequency-dependence in the ac magnetic susceptibility is found down to 50 mK, as shown in Fig.~\ref{fig:fig2}\textit{C}, indicating that there is no spin freezing present. The nature of the frequency-independent hump around 0.35 K is not clear, but a similar feature found in the Co-based material Na$_{3}$Co(CO$_{3}$)$_{2}$Cl at 17 K has been attributed to neither a magnetic ordering nor a spin-glass transition \cite{Fu2013}. 

\begin{figure*}[t]
\centering
\includegraphics[width=12cm,height=12.0cm]{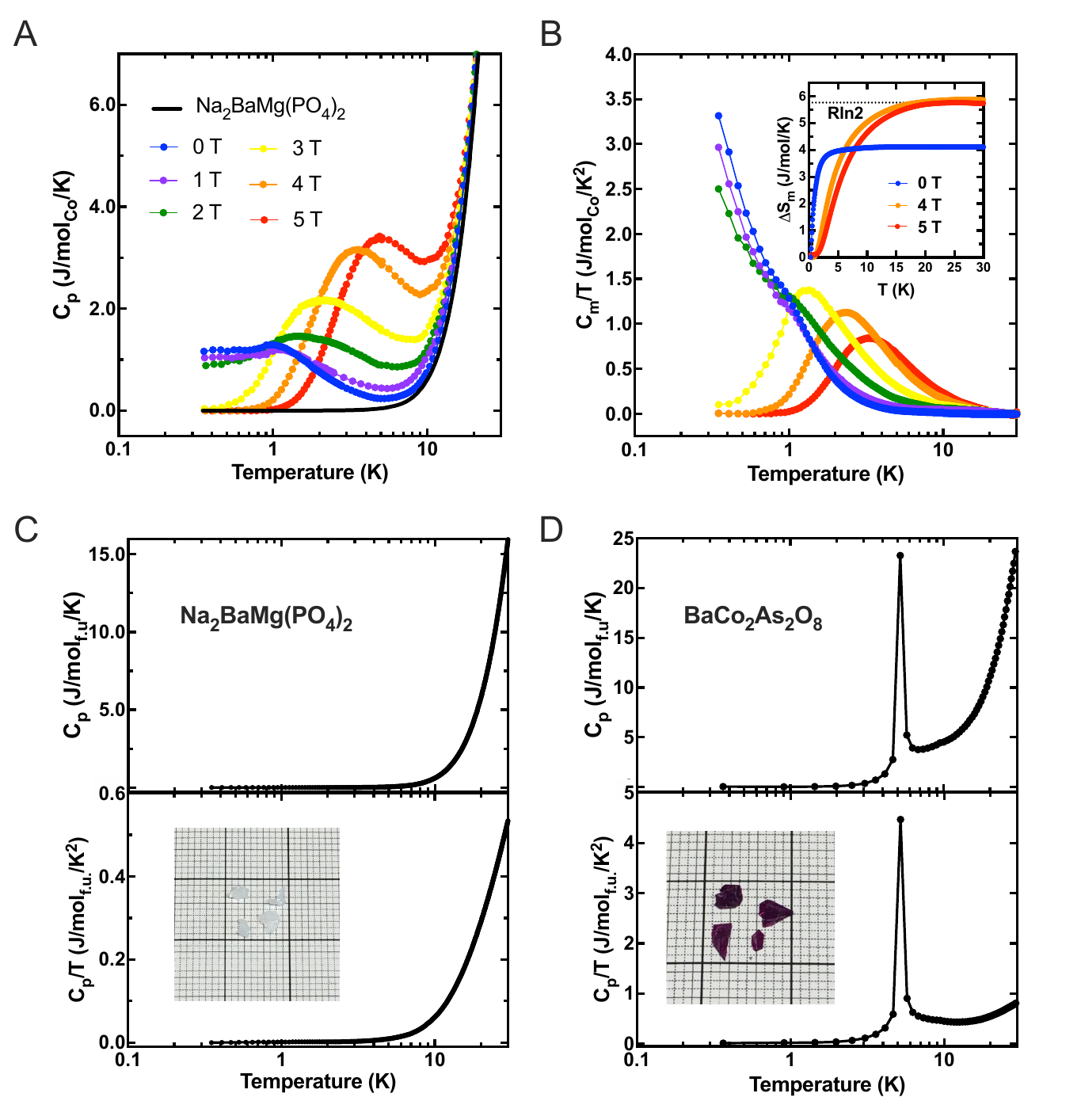}
\caption{Magnetic specific heat results for \nbcpo. \textit{(A)} Heat capacity $C_{p}$ as a function of temperature for a \nbcpo\ single crystal measured under several applied magnetic fields. \textit{(B)} $C_{m}/T$ as a function of temperature at several magnetic fields. Inset: Temperature dependences of integral magnetic entropy under magnetic fields of 0 T, 4T and 5 T for \nbcpo. \textit{(C, D)} Heat capacities for \nbmpo\ and BaCo$_{2}$As$_{2}$O$_{8}$, respectively, measured in the same temperature range. }\label{fig:fig3}
\end{figure*}

In Fig.~\ref{fig:fig3}\textit{A} we present the specific heat $C_{p}$ down to 0.35~K for a \nbcpo\ single crystal. Under zero field, neither a peak, as would be expected for a well-defined magnetic phase transition, nor a broad hump, as would be typical for a spin glass system \cite{Ma2018, Li2015, Zhong2018} is observed, consistent with all our other data. For applied field above 3 T, there is still no evidence of long-range magnetic ordering but instead a broad peak appears. This broad hump is associated with Schottky anomaly that shifts to higher temperatures at higher fields. 

A dramatic transition in behavior is found at a critical field, denoted as H$^{\ast}$, around 2$\sim$3 T. A single crystal of the nonmagnetic isostructural analogue \nbmpo\ was used for subtraction of phonon contribution to the total heat capacity, allowing us to calculate the magnetic specific heat Cm for \nbcpo, as shown in Fig.~\ref{fig:fig3}\textit{B}. The upturning feature that appears at the lowest temperatures in $C_{m}/T$ for \nbcpo\ is unexpected. A similar upturning feature found in some materials \cite{Yamashita2008, Collan1970, Yamashita2011} has been attributed to a nuclear Schottky contribution arising from the nuclei of the constituent atoms. In our case, however, we rule out this possibility by separately measuring the specific heats of the related materials \nbmpo\ (Fig.~\ref{fig:fig3}\textit{C}) and BaCo$_{2}$As$_{2}$O$_{8}$ down to 0.35 K (Fig.~\ref{fig:fig3}\textit{D}). These materials contain all the atoms (and their nuclei) that are present in \nbcpo\ but do not show this feature in the studied temperature range. In addition, the magnitude of the measured magnetic specific heat for \nbcpo\ in this temperature range is several orders of magnitude larger than that of a nuclear Schottky contribution \cite{Yamashita2008, Collan1970, Yamashita2011}, which further indicates that there is a tremendous amount of magnetic entropy present at low temperatures in this new material. To determine the magnetic entropy $\Delta S_{m}$, we first interpolated the curve of $C_{m}/T$ vs. $T$ to 0~K and then performed the integration. Due to the upturning feature at the lowest temperature, the integration for magnetic entropy is not complete. In fact, for a $S$=1/2 QSL system, the expected total magnetic entropy should reach to exactly $Rln2$ ($R$ = the ideal gas constant). Nevertheless, the obtained magnetic entropy for zero field reaches only 0.71$Rln2$ by 30 K. The difference between this value and $Rln2$ indicates that more magnetic entropy is present below 0.3 K. By applying a field larger than 4 T, conventional thermodynamic behavior returns, giving rise to a total entropy that is precisely $Rln2$ (inset of Fig. 3$B$, selected field are presented for clarity).

\begin{figure*}[t!]
\centering
\includegraphics[width=14cm,height=10.0cm]{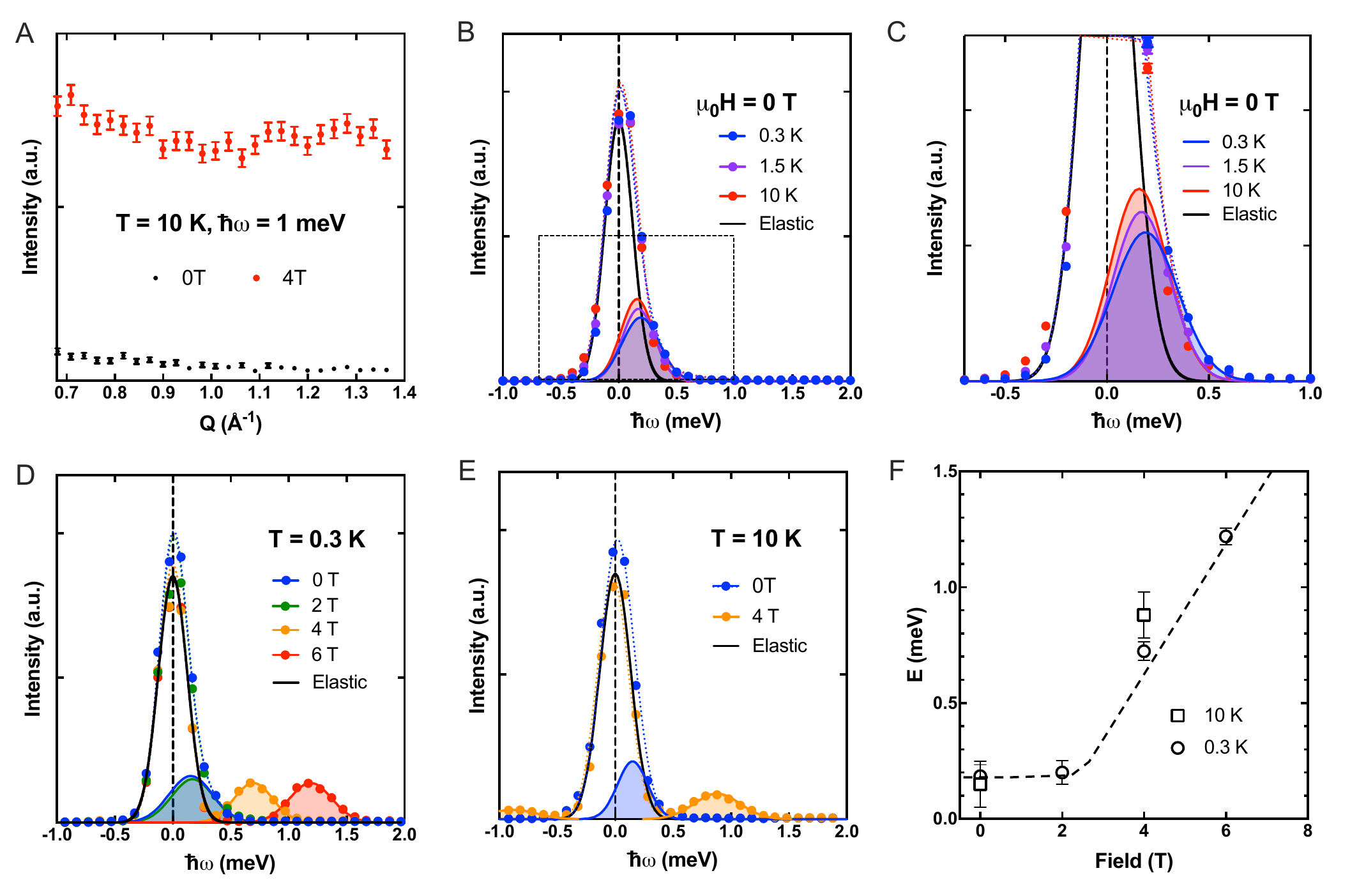}
\caption{Evidence for strong quantum fluctuations in \nbcpo. \textit{(A)} Inelastic neutron scattering intensity ($\hbar \omega$ = 1 meV) as a function of wave vector $Q$ under magnetic fields of 0 and 4 T measured at 10 K. \textit{(B)} Zero-field quasi-elastic neutron scattering spectra measured at 0.3 K, 1.5 K and 10 K, respectively. The dotted box in panel \textit{(B)}  is expanded in \textit{(C)}. \textit{(C)} Detailed view of the neutron scattering spectra showing the asymmetric characteristics at all temperatures. \textit{(D, E)} Energy dependence of the scattering for \nbcpo\ with applied magnetic fields measured at 0.3 K and 10 K, respectively. In \textit{(B-E)}, dots are experimental data, the dashed lines are the fits described in the text, the black curves indicate the fits to the incoherent elastic scattering, the colored solid lines as well as their shaded areas show the contribution from spin fluctuations under different conditions. \textit{(F)} Energy of characteristic magnetic excitations under applied magnetic fields ranging from 0 to 6 T. Dashed lines in \textit{(F)} are guides to the eyes. \label{fig:fig4}}
\end{figure*}

Characteristic signatures for strong quantum fluctuations are observed in inelastic neutron scattering measurements on the powder sample. Low energy (1 meV) scattering intensities measured with or without an applied magnetic field at 10 K are shown in Fig.~\ref{fig:fig4}\textit{A}. We see similar intensities over the $Q$ range we measured and the magnetic excitation intensities show no significant $Q$ dependence. Having a relatively flat $Q$-dependence implies that these intensities are related to fluctuations localized in real space, possibly due to strong frustration in the system. While these intensities do respond to an external magnetic field, it is evident that these are indeed localized spin fluctuations. 

We now focus on the temperature and field dependence of the magnetic excitations seen by quasielastic neutron scattering at an arbitrary position in scattering space, $Q$ = 1.276 {\AA}$^{-1}$, corresponding to (H, K, L) = (1/3, 2/3, 0) in reciprocal space. Fig.~\ref{fig:fig4}\textit{B} shows the energy dependence of the scattering intensity obtained at temperatures ranging from 0.3 to 10 K. The overall shape and intensity do not change much with temperature, indicating that the low-energy spin fluctuations are not much affected by temperature up to 10 K. This observation coincides with that of $\alpha$-RuCl$_{3}$\cite{Banerjee2016}, which shows an insignificantly affected intensity at higher temperature for the higher-energy mode accounted for proximate Kitaev QSL state. Moreover, one can see the “quasi-elastic” peak is not actually symmetric about $\hbar \omega$ = 0. The degree of asymmetry slightly decreases when heating to 10 K, as shown in the zoomed-in plots in Fig.~\ref{fig:fig4}\textit{C}, which indicates that the spin fluctuations slightly shift to slightly lower energy. The measured spectra, indicated by the dashed lines, can be fit as a sum of incoherent elastic scattering (black line) and a quasi-elastic spin excitation (colored lines) in Fig.~\ref{fig:fig4}\textit{B}. After subtraction of the incoherent scattering, one can determine a low-energy magnetic excitation (solid colored lines) with energy scale ~0.18-0.2 meV. 


The energy resolution of the SPINS instrument is not enough to cleanly separate the inelastic magnetic excitations from the elastic background. Nonetheless, the magnetic field response of these low-energy spin excitations at both 0.3 K and 10 K reflects the inelastic nature of the scattering. At 0.3 K (Fig.~\ref{fig:fig4}\textit{D}), the spectrum is dominated by excitations from the ground state to the excited state and thus we only see intensities on the neutron energy loss side (positive energy side); while at 10 K (Fig.~\ref{fig:fig4}\textit{E}), the excited state becomes populated and the spectrum also includes intensities coming from the energy gain side (negative energy side). The relative intensities follow the detailed balance principle $S(Q, -\hbar \omega)$ =$e^{-\hbar \omega/k_{B}T}S(Q, \hbar \omega)$. The splitting observed at high field is likely of a conventional Zeeman type. Note that at higher fields in both temperatures, when the quasi-elastic spin excitation peak shifts away from the elastic background to higher energy, the remaining elastic feature (black line) become symmetric around 0, showing in Fig.~\ref{fig:fig4}\textit{D, E}. This further indicates that the asymmetric peaks observed at low fields are indeed due to extra magnetic components on the positive energy side. The energy needed in order to excite such magnetic states at different fields can be extracted from the fitted peak center, as plotted in Fig.~\ref{fig:fig4}\textit{F}. At a field smaller than H$^{\ast}$ ($\mu_{0}H$ = 0T, 2T), the excitation energy is small and relatively constant, seeming to indicate a constant energy gap between the ground state and the spin-excitation continuum. In contrast to the field response of the spin-wave excitations in a conventional magnet, the preserved spin excitations observed here further suggest that we have an unconventional magnetic system. In support of this, the unaffected spinon excitations under weak field have been discussed in earlier reports, both experimentally \cite{Shen2018} and theoretically \cite{LiYD2017}. With a strong field larger than H$^{\ast}$ ($\mu_{0}H$ = 4T, 6T), the magnetic excitations are now peaked at higher energies, clearly separated from the incoherent elastic peak, and the excitation energy increases linearly with field. The linear relation in the high field regime seems to indicate a conventional Zeeman splitting behavior of magnon excitations, agrees with the expectation for the nearly polarized magnetic state. 

Interestingly, the exotic nonlinear field response agrees well with the specific heat measurements. At fields less than H$^{\ast}$, one still observes a non-zero specific heat response at the lowest temperature. It is reasonable to expect that with no applied magnetic field or small field, the quantum spin fluctuations are "forced" into the low energy (< 0.2 meV) channels, and therefore contribute to the anomalous specific heat in this quantum spin liquid. Only with large field when the conventional Zeeman splitting behavior dominates, these spin excitations can be shifted to higher energies, and the specific heat of the system reverts back to normal behavior.

From the energy-integrated intensity of the spin excitations seen in the neutron scattering data, one can perform an absolute normalization to determine the size of spin fluctuations \cite{Xu2013}. The calculated fluctuating moment is then $m = g\sqrt{S(S+1)}$ = 1.52$\mu_{B}$/Co$^{2+}$ indicating strong magnetic fluctuations compared to the total local moment. This also provides an independent estimation of the effective magnetic moment $S$ = 1/2. We note that the integrated intensities of spin excitations at different temperatures and fields are comparable, indicating that most of the spin fluctuation spectral weight is confined to low energy, and shifts to higher energies at sufficiently high applied field. 

\section*{Conclusions}
Our observations on the new compound \nbcpo\ reveal its potential as a new quantum spin liquid candidate. No magnetic ordering or spin freezing is observed down to 0.05 K. Thermodynamic measurements show that there is tremendous magnetic entropy present below 1 K, a reflection of the presence of a very large spin density of states at low temperatures. Inelastic neutron scattering measurements show a flat $Q$-dependence for the low-energy magnetic excitations, corresponding to localized low-energy spin fluctuations, which are persistent with changing temperature and weak magnetic field. The exotic temperature and magnetic field response of these low-energy magnetic excitations further indicates the nontrivial nature of the magnetic ground state of \nbcpo. Since this newly discovered compound is structurally superior to the other existing QSL candidates, any broad continuum excitation found in future inelastic neutron experiments could be associated with the spinons, which are a consequence of the QSL state, instead of disorder.

\section*{Materials and methods}

\subsection*{Materials synthesis}
Single crystals of \nbcpo\ for dc susceptibility, specific heat and thermal conductivity measurements were synthesized by the flux method. Dried NaCl (99\%, Alfa Aesar), BaCO$_{3}$ (99.95\%, Alfa Aesar), CoO (99\%, Alfa Aesar), and (NH$_{4}$)$_{2}$HPO$_{4}$ (99.5\%, Sigma-Aldrich) were mixed stoichiometrically and ground well with the flux media NaCl in a molar ratio of 1:5. The mixed starting materials were loaded in an alumina crucible with lid, and then heated up to 950 \degrees\ for 2 hours, followed by a slow cooling procedure of 3 \degrees\ /hr to 750 \degrees. The obtained pink single crystals were manually separated from the bulk. Polycrystalline \nbcpo\ samples for ac susceptibility and neutron scattering were prepared by a solid-state reaction: a stoichiometric mixture of the above starting materials was ground together with a catalyst NH$_{4}$Cl (1:0.5 in molar ration), sintered in air, in stages, to 700 \degrees\ for 24 hrs. 

\subsection*{Single-crystal X-ray diffraction measurements and refinements}
The crystal structure of the title compound was determined by single crystal X-ray Diffraction. The diffraction data were collected at 298(2) K with a Kappa Apex2 CCD diffractometer (Bruker) using graphite-monochromated Mo-K$\alpha$ radiation ($\lambda$ = 0.71073 $\AA$). The raw data were corrected for background, polarization, and the Lorentz factor and multi-scan absorption corrections were applied. Finally, the structure was analyzed by the Intrinsic Phasing method provided by the ShelXL structure solution program \cite{Sheldrick2015} and refined using the ShelXL least-squares refinement package with the Olex2 program \cite{Dolomanov2009}. The ADDSYM algorithm in program PLATON was used to double check for possible higher symmetry \cite{Spek2003}.

\subsection*{Magnetization and thermodynamic measurements}
The \textit{dc} magnetic susceptibility was measured in a physical property measurement system (PPMS) that cooled to 1.8 K (PPMS-DynaCool, Quantum Design). The \textit{ac} susceptibility measurements on a polycrystalline pellet sample were conducted in a Quantum Design PPMS DynaCool with a dilution refrigerator insert (ACDR measurement option) allowing for a measurement down to 0.05 K. Specific heat was measured in a PPMS equipped with a $^{3}$He inset. All heat capacity measurements were carried out with external magnetic field applied parallel to the c axis of a single crystal. 

\subsection*{Neutron-scattering measurements} 
Elastic and inelastic neutron scattering experiments were performed on the NG5-Spin Polarized Inelastic Neutron Spectrometer (SPINS), a cold neutron source located at the NIST center for Neutron Research. The 25-g powder sample was tightly packed into an aluminum sample holder, which was then mounted in a closed-cycle refrigerator (CCR) equipped with a $^{3}$He insert for cooling. Before the measurements at a particular temperature, we held the system at temperature overnight to guarantee a homogeneous temperature of our powder sample. The collimators used for this experiment were Guide-80-80-open, with a fixed incident energy of 5 meV ($\lambda$=4.045 $\AA$). Energy resolution was about 0.3 meV under current settings. The wave vector Q is expressed as $\lvert Q\rvert$ =2$\lvert k_{i} \rvert sin2\theta$, where $2\theta$ is the angle between the incident and the final beam. 

\subsection*{Data Availability} 
The crystal structure information is available in the CSD database, number 1866950. The physical property and neutron scattering datasets generated during and/or analyzed during the current study are available from the corresponding author on reasonable request.

\begin{acknowledgments}
We acknowledge the Applications Group at Quantum Design for measuring the ac susceptibility between 0.05 and 0.35 K. We are thankful to the support of the National Institute of Standards and Technology, U.S. Department of Commerce, for providing the neutron research facilities used in this work. We thank N. P. Ong and J. Tranquada for very helpful discussions and communications. The materials synthesis and magnetic characterization was supported by the Gordon and Betty Moore EPiQS program, grant GBMF-4412. The crystal structure determination was supported by U.S. Department of Energy, Office of Science, EFRC grant number DE-SC0019331. 
\end{acknowledgments}

\begin{table}
\centering
\caption{Crystal data and ambient temperature crystal structure refinements for \nbcpo.}
\begin{tabular}{lr}
\hline
\hline
Formula & \nbcpo   \\
Formula mass ($g/mol$) & 432.21   \\
Crystal system & trigonal   \\
Space group, Z & P$\bar{3}$m1 (No.164), 1  \\
$a$ ({\AA}) & 5.3185(1) \\
$c$ ({\AA}) & 7.0081(1) \\
$V$ ({\AA}$^{3}$) & 171.676(5) \\
$T$ (K) & 298 (2) \\ 
$\rho$ (cal) ($g/cm^{3}$) & 4.180 \\
$\lambda$ ({\AA}) & 0.71073 \\
F (000) & 199 \\
$\theta$ (deg) & 2.91-33.14 \\
Crystal size ($mm^{3}$) & 0.065 $\times$ 0.065 $\times$ 0.04 \\
$\mu$ ($mm^{-1}$) & 8.721 \\
Final R indices & R$_{1}$ = 0.0260, $\omega$R$_{2}$ =0.0574 \\
R indices (all data) & R$_{1}$ = 0.0272, $\omega$R$_{2}$ =0.0580 \\
Goodness of fit & 1.197 \\
\hline
\hline
\end{tabular}
\end{table}

\begin{table}
\caption{Wyckoff positions, coordinates, occupancies, and equivalent isotropic displacement parameters for \nbcpo.}
\begin{tabular}{ l r r r r r r}
\hline
\hline
Atom & \shortstack{Wyckoff \\ site} & x & y & z & S.O.F. & U$_{eq}$ \\
\hline
Ba1 & 1a & 0 & 0 & 0 & 1 & 0.01031(18) \\
Co1 & 1b & 0 & 0 & 0.5 & 1 & 0.0069(3) \\
P1 & 2d & 0.3333 & 0.6667 & 0.2420(3) & 1 & 0.0064(4) \\
Na1 & 2d & 0.3333 & 0.6667 & 0.6794(6) & 1 & 0.0162(8) \\
O2 & 2d & 0.3333 & 0.6667 & 0.0250(8) & 1 & 0.0128(12) \\
O1 & 6i & 0.1793(5) & 0.8207(5) & 0.3188(6) & 1 & 0.0417(14) \\
\hline
\hline
\end{tabular}
\end{table}

\end{document}